\documentclass[12pt]{iopart}
\usepackage{braket}

\usepackage{graphicx}
\usepackage{amsfonts}
\usepackage{amssymb}
\usepackage{amsbsy}
\usepackage{latexsym}

\newcommand{\none}{\nonumber\\}

\def\be {\begin{equation}}
\def\ee  {\end{equation}}
\def\bea {\begin{eqnarray}}
\def\eea {\end{eqnarray}}
\def\nn {\nonumber}
\newcommand{\half}{{\frac{1}{2}}}

\begin{document}

\title{Bosonic string theory with dust}

\author{Jack Gegenberg and Viqar Husain}

\address{Department of Mathematics and Statistics and Department of Physics\\University of New Brunswick\\
Fredericton, NB E3B 5A3, Canada}



\begin{abstract}

We study a modified bosonic string theory that has a pressureless  ``dust'' field on the string worldsheet.  The dust is a real scalar field with unit gradient which breaks conformal invariance.  Hamiltonian analysis  reveals  a  time reparametrization constraint  linear in the dust field momentum and a  spatial diffeomorphism constraint. This feature provides  a natural ``dust time" gauge in analogy with the parametrized particle. In this gauge we give a  Fock quantization of the theory, which is complete and self-consistent in $d<26$. The Hamiltonian of the theory is not a constraint; as a consequence  the Hilbert space and mass spectrum are characterized by an additional parameter, and includes the usual string spectrum as a special case. The other sectors provide new particle spectra, some of which do not have tachyons. 
\end{abstract}

%

\section{Introduction}

String theory is a promising program for reconciling gravity and elementary particle physics.  The classical string theory is  a conformal and diffeomorphism invariant 2-dimensional field theory.  Preservation of these invariances in the quantum theory requires that strings propagate in a target spacetime that is  either 26 dimensional for the bosonic string, or 10  for the superstring.  This is both a strength and weakness of string theory;   higher dimensionality allows the possibility of  Kaluza-Klein type unification of the fundamental interactions,  but  at the same time poses the  challenge of observing extra dimensions, which could be any of a vast range of possible Calabi-Yau manifolds.

In this paper we construct a modification of bosonic string theory that is anomaly free in any dimension.  This is done by adding a new world sheet field to the Polyakov action, while keeping the metric and embedding fields unaltered.  The new field cannot be absorbed as an extra embedding coordinate due to the imposed condition that the norm of its gradient is fixed.  This  feature breaks conformal invariance at the classical level.

The approach we take uses an idea espoused for four-dimenisonal gravity in \cite{hp}, where it is shown that adding such a field to the action of general relativity  leads to a Hamiltonian constraint  that  is linear in the dust momentum.  This enables a simple algebraic solution of the classical Hamiltonian constraint to give a physical Hamiltonian if the dust field is chosen as a time gauge. It  leads to a theory that can in principle be quantized non-perturbatively.  It is therefore of interest to see what a similar construction yields for string theory. In the next two sections we describe  the covariant and Hamiltonian formulations of this modified bosonic string moving in an arbitrary dimensional Minkowski spacetime, including a comparison with the usual string. In Section IV we give the canonical quantization of the time gauge fixed theory in the  Fock Hilbert space, derive the mass spectrum formula, and show that negative norm states are eliminated in manner similar to that in the standard string.  The last section contains a summary and discussion.

\section{Classical Action}

The action consists of two terms,  the Polyakov  action of the bosonic string in d dimensions and that  of a pressureless irrotational dust.  The dust and the string embedding variables couple to the worldsheet metric $g_{ij}$, but not directly to each other:
\be
S[X,g,T,M]=S_B[X,g]+S_D[T,M,g],
\ee
where
\bea
S_B[X,g]&=&-\frac{1}{2\kappa}\int_\Sigma d^2x\sqrt{-g}g^{ij}X^A_{,i}X^B_{,j}\eta_{AB};\\
S_D[T,M,g]&=&-\int_\Sigma d^2x\sqrt{-g}M(g^{ij}T_{,i}T_{,j}+1).
\eea
Here the $x^i=t,\sigma$ are coordinates on the worldsheet $\Sigma$ and the $X^A(t,\sigma), A=0,1,...d-1$ are the embedding coordinates of the string worldsheet in the target Minkowski spacetime $M_d$ with flat metric $\eta^{AB}=$diagonal $[-1,1,...,1]$.  The worldsheet manifold is the product $\Sigma = S\times \mathbb{R}$.    The gradient $T_{,i}(t,\sigma)$ of the worldsheet scalar field $T(t,\sigma)$ gives the flow lines of the dust field and is constrained to be timelike.  Finally, $M(t,x)$ is its mass density of the dust field.  The string coupling constant is $\kappa:=2\pi\alpha'$.

 We note that (i) $S_D$ is {\it not invariant under conformal transformations} due to the term $M\sqrt{-g}$ in it, (ii)  the fields $T$ and $M$ do not have the geometrical role of the world sheet embedding fields $X^A$, because embedding fields do have a fixed norm condition, unlike $T$; although the kinetic part of $T$ can be absorbed  into the Polyakov form by extending the target space metric from diag$[-1, 1, \cdots 1]$ to diag$[-1,1, \cdots 1, M]$, the conformal symmetry breaking term  $\int d^2\sigma \sqrt{-g} M$ remains as a separate term in the action.

Parametrizing the world sheet metric using the lapse and shift functions as
\cite{henneaux}
 \be
ds^2= e^{2\rho}h_{ij}(N,N^\sigma)dx^i dx^j \equiv e^{2\rho}\left[-N^2 dt^2+(d\sigma+N^\sigma dt)^2\right]. \label{wsheetg}
\ee
the action becomes
\be
S= S_B[X,h] - \int d^2x \sqrt{-h} M\left(h^{ij}T_{,i}T_{,j} + e^{2\rho}\right). \label{action2}
\ee
This shows that  a conformal rescaling of the world sheet metric  changes the norm of the gradient of the dust field.

Now as part of the definition of the theory, we specify the conformal factor $e^{2\rho}$ as a fixed background structure. That is, the variational principle is specified by varying all fields in the action except $\rho$. Now we can work with an arbitrary but fixed $\rho$, but for technical convenience we set $\rho(t,\sigma)=0$ in $S_D$ {\it as the definition of the theory}.\footnote{It is possible to  leave the function $\rho(t,\sigma)$ as an arbitrary but fixed  ``background'' field in the subsequent analysis.} This construction is very much akin to the choice of a Minkowski spacetime background in quantum field theory on fixed background.

The fields to be varied in the action are  therefore $X^A$, $T$, $M$, $N$ and $N^\sigma$. If the theory were defined by varying $\rho$ as well, its equation of motion would give $M=0$, and it would reduce to the usual bosonic string; fixing $\rho$ as a partial ``background'' also separates conformal invariance from spacetime diffeomorphism invariance, a feature that will be apparent in the canonical theory described below.

With this prescription the  equations of motion for $X^A$, $M$ and $T$ that  follow from (\ref{action2}) (with $\rho=0$) are respectively
\bea
&\Box_h X^A=0,\label{eq:X}\\
&h^{ij}T_{,i}T_{,j} +1=0,\label{eq:M}\\
&\nabla_i \left(\sqrt{-h} h^{ij}MT_{,j} \right)=0, \label{eq:T}
\eea
where   $\nabla$ is the covariant derivative of $h_{ij}(N,N^\sigma)$,   and $\Box_h :=\nabla^i\nabla_i$.

These equations are supplemented by two additional ones obtained by varying the action with respect to $N$ and $N^\sigma$. However, in the covariant formulation these equations are messy and reveal little about the underlying theory. Therefore we write them out in the Hamiltonian theory in the next section, where their form is simple and interpretation is manifest.

\section{Hamiltonian formulation}

The canonical form of the action is obtained by substituting the metric (\ref{wsheetg}) with $\rho=0$ into the covariant action, and replacing all time derivatives of fields by their corresponding momenta. We use the notation  $\dot M:=\partial_t M,M':=\partial_\sigma M$.   Since the action does not contain time derivatives of the metric, the momenta conjugate to the metric functions $N$ and $N^\sigma$ vanish:
\be
  P_N=0,\ \ \     P_{N^\sigma}=0.
\ee
The momenta conjugate to the variables $X^A,T,M$ are
\bea
P_A&=&\frac{1}{\kappa N}\left(\dot X^B-N^\sigma X^B{}'\right)\eta_{AB},\label{Pf}\\
P_T&=&\frac{2 M}{N}(\dot T-N^\sigma T'), \label{PT}\\
P_M&=&0.
\eea
Therefore the action (\ref{action2}) (with  $\rho=0$) becomes
\be
S = \int dtd\sigma\ \left( P_T\dot{T} + P_A\dot{X}^A - N{\cal H} - N^\sigma {\cal H}_1   \right)  \label{canA}
\ee
 where
 \bea
 {\cal H} &=&
\frac{P_T^2}{4M} +M\left(1+ (T')^2\right)+\half\left(\kappa (P_A)^2+\kappa^{-1}(X^A{}')^2\right),\label{eq:G}\\
{\cal H}_1&=& T'P_T  - X^A{}'P_A.\label{eq:F}
\eea
At this stage the canonical action is a functional of the canonical pairs $(X^A, P_A)$, $(T, P_T)$, $M$, and the lapse and  shift functions $N$ and $N^\sigma$.

It is now clear that the variation of the action with respect to $N$ and $N^\sigma$ give the diffeomorphism and Hamiltonian constraints
\be
{\cal H}_1 =0, \ \ \ \ \ {\cal H} =0.
\ee
We can eliminate $M$ by using its equation of motion and substituting it back into the Hamiltonian density  (\ref{eq:G}). Varying with respect to $M$ gives
\be
M^2 =  \frac{P_T^2}{4(1 + T'^2)} .  \label{Meqn}
\ee
 With this  the Hamiltonian constraint  (\ref{eq:G})  may be written as
\be
{\cal H} =  sgn(M) P_T \sqrt{1+ T'^2}   +   \half\left[\kappa (P_A)^2+\kappa^{-1}(X^A{}')^2\right] =0. \label{Hc}
\ee
This modified string theory has  $d+1$   local configuration degrees of freedom  ($d$ fields $X^A$, and $T$), which are subject to  two local first class constraints for a net $d-1$ local degrees of freedom; for the standard bosonic string, a similar count gives  $d-2$.  It is this fact which ultimately leads to different and richer particle spectrum in the quantum theory.

\subsection{Dust time gauge}

In the conventional bosonic string, a standard procedure is to impose the light cone gauge, where the world sheet time $t= (X^0+X^1)/\sqrt{2}$. This breaks target space Poincare invariance. It must then be checked that this symmetry is unbroken at the  quantum level by computing the algebra of its generators. As is well-known the algebra closes only in 26 dimensions.

 We choose the gauge $T(\sigma,t) =t$. This is natural for two reasons:  it maintains target space Poincare invariance, and leads to a simple form of the physical Hamiltonian $H_P=-P_T$,  as we now show.

It is evident that the Hamiltonian constraint (\ref{Hc}) and the gauge condition $T(\sigma,t)-t =0$ are a second class pair; their Poisson bracket is
\be
\{T(\sigma,t), {\cal H}(\sigma',t)  \} = \sqrt{1 + (T')^2}\ \delta(\sigma-\sigma'),
\ee
where the right hand side is obviously not one of the constraints. Therefore we must either work with Dirac brackets, or solve the constraints explicitly. The latter is an easy option since the Hamiltonian constraint  is readily solved to give $P_T$ as a function of  $X^A$ and $P_A$. Substituting $T=t$ and $T' =0$ into the constraints gives  \be
sgn(M) P_T = - \frac{1}{2}\   \left[\kappa (P_A)^2+\kappa^{-1}(X^A{}')^2\right], \\\label{eq:G1}
\ee
and the diffeomorphism constraint simplifies to
\be
{\cal H}_1=P_A X^A{}' =0\label{diff0}.
\ee

As with any canonical gauge fixing,  we require that the gauge $T(\sigma,t)=t$  be preserved under time evolution. This gives conditions on the lapse and shift functions $N$ and $N^\sigma$.
The consistency condition is
\be
 \dot {T}(\sigma,t) = \{ T(\sigma,t), \int d\sigma \left(N{\cal H} +  N^\sigma {\cal H}_1 \right) \} =1,
\ee
where the Poisson bracket much be computed first, and then the gauge condition inserted. This gives
\be
\dot{T} =  \left(N\sqrt{1+(T')^2}  +    N^\sigma T'\right)|_{T=t}  = N,
\ee
which implies $N =1$ (since $\dot{T}=1$), and leaves $N^\sigma$ unconstrained.

We now notice that eqn. (\ref{PT}) becomes $P_T = 2M$ \cite{JSdust}, so that in eqn. (\ref{eq:G1}) $sgn(M)P_T = P_T$.  Therefore the (time) gauge fixed action is
\be
S^{GF} = \int dt d\sigma \left( P_A\dot{X}^A  + P_T  -N^\sigma {\cal H}_1\right). \label{S-gf}
\ee
From this we can identify the physical Hamiltonian density, up to the constraint term $N^\sigma {\cal H}_1$ as,
\be
{\cal H}_P :=  -P_T = \frac{1}{2}  \left[\kappa (P_A)^2+\kappa^{-1}(X^A{}')^2\right] .\label{H-phys0}
\ee
where the last equality follows from solving the Hamiltonian constraint.

We note that the time gauge fixed  action  (\ref{S-gf})  retains target space Poincare invariance due to the fact that the time gauge condition $T=t$ does not break this  invariance (unlike lightcone gauge in the standard bosonic string). Furthermore it is  the canonical action of $d$ scalar fields in two spacetime dimensions with an additional local symmetry --  spatial diffeomorphisms. (The other symmetry is the afore mentioned global target space Poincare invariance.)

This completes the classical description of the theory.\footnote{The features discussed are not unique to two world sheet dimensions -- dust fields may be introduced for membranes etc.  and the structure of the theory is unchanged, ie. the true Hamiltonian is that of free scalar fields with local spatial diffeomorphism invariance of the world sheet.} In the following we restrict attention to the closed string, so that  this symmetry is local Diff($S^1$) invariance.

\subsection{Comparison with the bosonic string}

At this stage  it is useful to make a comparison with the constraint structure of the standard closed bosonic string. This  further highlights the differences between the two theories, and gives  a helpful comparison  of their descriptions in terms of  oscillator variables.

The usual closed string constraints in the Hamiltonian theory are  \cite{henneaux}:
\bea
 && \frac{1}{2}  \left[\kappa (P_A)^2+\kappa^{-1}(X^A{}')^2\right] =0, \nn\\
  &&P_A X^A{}' =  0.
\eea
This shows the basic  difference between the two theories:  the  function in the first constraint {\it is} the non-vanishing physical Hamiltonian density  (\ref{H-phys0}) of the dusty string. As elaborated above this difference arises because the Hamiltonian constraint of  the latter theory is solved (in the $T=t$ gauge) to yield the physical hamiltonian, which turns out to be the same functional form, ie. the Hamiltonian function of $d$ free scalar fields.

The  Virasoro generators of the closed string are defined as the Fourier modes of linear combinations of these constraints \cite{henneaux}:
\bea
L_n &:=&  \frac{1}{2}\int_0^{2\pi}  d\sigma\  e^{in\sigma} \left( {\cal H}_P + {\cal H}_1  \right), \\
\bar{L}_n &:=&  \frac{1}{2}\int_0^{2\pi}  d\sigma\  e^{-in\sigma} \left( {\cal H}_P - {\cal H}_1  \right).
\eea
From these definitions  we have $L_n^* = L_{-n}$ and $\bar{L}_n^* = \bar{L}_{-n} $. The physical Hamiltonian ${\cal H}_P$ (\ref{H-phys0}) and constraint ${\cal H}_1$ (\ref{diff0}) of the theory with dust may be written in terms of these generators as
 \bea
  \int_0^{2\pi} d\sigma \  e^{in\sigma} {\cal H}_P &=& L_n + \bar{L}_n^*,  \\
  \int_0^{2\pi} d\sigma \  e^{in\sigma} {\cal H}_1 &=& L_n - \bar{L}_n^*.  \label{Ln}
 \eea
 We remind the reader that unlike the  standard string, the first equation is no longer a constraint in the theory with dust --  the physical Hamiltonian in dust time gauge is the non-vanishing expression
\be
H_P := \int_0^{2\pi} d\sigma\  {\cal H}_P = L_0 + \bar L_0,
\ee
with the diffeomorphism contraints
\be
  L_n = \bar L_n^*,  \ \ \ \ \  \forall\  n. \label{diffeo-osc}
\ee
This Hamiltonian and set of constraints can be written in terms of oscillator variables in the standard way  by  substituting the  spatial Fourier expansions \cite{henneaux}
\bea
X^A(\sigma,t)&=&x^A_{c}(t) + \sqrt{\frac{\alpha'}{2}} \sum_{n>0} n^{-\half}\left[ c^A_n(t) e^{-in\sigma}+\bar c^A_n(t) e^{in\sigma}  + {\rm c.c.}\right] \\
P^A(\sigma,t)&=& \frac{p^A_{c}(t)}{2\pi} + \frac{1}{2\pi\sqrt{2\alpha'}} \sum_{n>0} n^\half\left[ -ic^A_n(t) e^{-in\sigma}-i\bar c^A_n(t) e^{in\sigma} + {\rm c.c.}\right]
\eea
 into the expressions for ${\cal H}_P$ and ${\cal H}_1$. The Poisson brackets of the coefficients in this mode expansion induced by the fundamental Poisson brackets of $X^A$ and $P_A$ are
 \be
\left\{x^A_{c},p_{lB}\right\} =\delta^B_C,\ \ \
\left\{c^A_m,c^{B*}_n\right\} =-i\eta^{AB}\delta_{mn},\ \ \
\left\{\bar c^{A}_m,\bar c^{B*}_n\right\}=-i\eta^{AB}\delta_{mn}.\label{eq:pbs}
\ee
The mode expansion gives
\be
H_P = \int_0^{2\pi} d\sigma\ {\cal H}_P = \frac{\alpha'}{2} p_c^2+  \sum_{n>0}n
\left( c_{nA}^* c_n^A + \bar{c}_{nA}^* \bar{c}_n^A  \right).  \label{Hphys-osc}
\ee
 The oscillator expansion of the $L_n$ and $\bar{L}_n$ may be similarly written down, giving the
familiar expressions.

\subsection{Complete gauge fixing}

In this section we describe a fixing of the remaining spatial diffeomorphism symmetry Diff($S^1$), together with a solution of the  corresponding constraint. The result is an unconstrained theory of  true physical variables.

The gauge condition we make is  $X^0(\sigma,t)=\sigma$. This has the two-fold advantage of providing a nice solution of the diffeomorphism constraint, and at the same time removing negative norm states in the quantum theory.  It is evident that this condition is second class with the diffeomorphism constraint, and so is a viable choice.   However $X^0(\sigma,t) = X^0(\sigma+2\pi,t)$ for the closed string case we are considering, whereas the function $f(\sigma)=\sigma$ is not periodic. This issue is easily dealt with by replacing the function $\sigma$ by the Fourier series
\be
f(\sigma):=\pi-2\sum_{n>0}\frac{1}{n}\sin{n\sigma},\label{fsig}
\ee
which converges to $\sigma$ in the domain $[0,2\pi)$.  So effectively the gauge fixing condition  is
\be
X^0(\sigma,t)=\left\{\begin{array}{lcl}\sigma, && 0\leq \sigma < 2\pi\\
0, && \sigma=2\pi.\end{array}\right.
\label{gfX0}
\ee
which is periodic, but discontinuous at $\sigma=2\pi$.

Evolution of this gauge  is given by  the Hamilton equation
\bea
\dot{X}^0 &=& \{ X^0(\sigma,t), \int_0^{2\pi} d\sigma'\left( {\cal H}_P + N^\sigma \right)  \} = \kappa P_0 + N^\sigma(X^0)' \nn\\
 &=& \kappa P_0 +N^\sigma= \dot{\sigma}=0,
\eea
which fixes the shift function: $N^\sigma= -\kappa P_0$.

Now solving strongly the diffeomorphism constraint gives
\be
 P_0 = -P_a (X^a)', \ \ \ \ \ \ \ i=1\cdots d-1. \label{P0}
\ee
Thus fixing the gauge and solving the constraint in this manner   eliminates the fields $P_0(\sigma,t)$ and  $X^0(\sigma,t)$. The resulting physical degrees of freedom are the spatial embedding variables $X^a(\sigma,t)$ and  conjugate momenta $P_a(\sigma,t)$.

The physical Hamiltonian density in terms of these variables is
\be
{\cal H}_P=  \frac{\kappa}{2} \left(   - (P_aX^{a\prime})^2   + P_aP^a \right)  + \frac{1}{2\kappa} \left( -1 + X^{a\prime} X_a ' \right).    \label{H-phys}
\ee

It is instructive to write this Hamiltonian in the oscillator variables. To do this we  note that the solution (\ref{P0}) for $P_0$  may be written in terms of the restricted set of generators $L_n^{sp}$ and $\bar{L}_n^{sp}$  in which the target space indices run over only the spatial components.  Defining
\be
\ell_n^\pm \equiv L_n^{sp} \pm \bar{L}_n^{sp*},
\ee
and using the mode definition  (\ref{Ln}) we have
\bea
-P_0 &=& P_aX^a{}'\nn\\
&=&  \ell_0^- + \sum_{n>0} \left(\ell_n^- e^{-in\sigma} + \ell_n^{-*}e^{in\sigma}  \right).
\eea
Therefore
\be
H_P = \int_0^{2\pi}    {\cal H}_P\ d\sigma
=  -\frac{1}{4\pi\alpha'}+ \ell_0^+
 - 2\pi^2\alpha'\left( (\ell_0^-)^2 + 2 \sum_{n>0}\ell_n^- \ell_n^{-*}     \right), \label{Hred-osc}
\ee
where
\bea
\ell_0^+&=&\frac{\alpha'}{2}p_{c}^2+\sum_{n>0}n\left(c^*_{an}c^a_n+\bar c^*_{an}\bar
c^a_n\right), \\
\ell_0^-&=&\sum_{n>0}n\left(c^*_{kn}c^a_n-\bar c^*_{an}\bar
c^a_n\right), \\
\ell_n^-=& -i&\sqrt{\frac{\alpha'n}{2}}p_a\left( c^a_n+\bar c^{a*}_n\right)+\sum_{m>0}\sqrt{m(m+n)}
\left(c^*_{am} c^a_{m+n}-\bar c^{a*}_{m+n} \bar c_{a m}\right)\none
&-&\half\sum_{m=1}^{n-1}\sqrt{m(n-m)}\left(c_{am}c^a_{n-m}-\bar c^*_{am}c^{a*}_{n-m}\right).
\eea

The last term in this Hamiltonian introduces a quartic coupling between the left and right movers, and between these and the center of mass (cm) momentum of the string. Furthermore this term if of order $\alpha'$, the same as the cm energy in $l_0^+$.    Of particular note  is that the cm momentum and oscillator modes are independent variables, unlike in the standard string, where the corresponding $H_P$ is constrained to be zero; the latter is what  provides the familiar connection between mass and oscillation modes, which is absent with the dust field.

\subsection{Poincare generators}

 The generators of  target space Poincare  transformations are
\be
M_{AB} = \int_0^{2\pi} d\sigma \left(X_AP_B - X_BP_A  \right).
 \ee

The dust field and its conjugate momentum plays no role here since these are   world sheet fields. With the above time and coordinate gauge fixing, the spatial Poincare generators $M_{ij}$ do not change,  but the time-space generator becomes
\be
M_{0a} = \int_0^{2\pi} d\sigma \left[\sigma P_a   + X_a(P_b X^b{}')  \right].
\ee
It is straightforward to check that the Poisson bracket $\{ M_{0a},M_{0b} \}$ closes on $M_{ab}$ as it should.

\section{Quantum Theory}

Using the  Hamiltonian  formulation described above, the theory may be quantized after only the time gauge fixing, or alternatively, after fixing  both time and space gauges.   The first may be referred to as ``Dirac quantization,'' since one first class constraint (spatial diffeomorphisms) remains to be imposed as an operator condition defining physical states, while  the second  approach is ``reduced phase space quantization'' (since all constraints are solved  and gauges fixed at the classical level).

 In the following we describe the former approach.  This mirrors in many ways the covariant quantization of the bosonic string, and gives  an appealing mass spectrum formula. The second approach also offers a complete and consistent quantization of the theory, without negative norm states; what is not achieved however  is the hard problem of finding the spectrum of the full Hamiltonian, although a perturbation theory from the free part is possible.

The problem is to find operator realizations of the physical Hamiltonian and the spatial diffeomorphism constraint, and to solve the energy eigenvalue problem $\hat{H}_P|\psi \rangle = E |\psi\rangle$, with $H_P$ given in (\ref{Hphys-osc}), and impose a quantum version of the diffeomorphism symmetry (\ref{diffeo-osc}).  Given the oscillator decompositions of the Hamiltonian and constraint,  Fock quantization is a natural choice. \footnote{We note that an alternative path is the polymer quantization method, in which the treatment of the spatial diffeomorphisms by the group averaging method is natural. See e.g. \cite{ALS,hk,mah} We leave this for a separate investigation.}

The Hilbert space is  the  Fock space whose elements are generated by  the mode creation operators $\hat{c}_n^{A\dagger}$ and   $\hat{\bar c}_n^{A\dagger}$, and the center of mass (CM) operators $\hat{x}_c^A, \hat{p}_c^A$; a state in the Hilbert space is therefore of the form
\be
| \{n_L\}, \{n_R\}, k\rangle,
\ee
where $\{n_L\}$ and $\{n_R\},$ represent a collection of occupation numbers of left and right moving internal modes, and $k$ is the centre of mass momentum. The inner product on these states is the standard one, with  a ``delta function normalization'' of the center of mass sector. The Fock vacuum is  given by
\be
\hat{c}_n^{a}|0,0,0\rangle =0, \ \ \ \ \  \hat{\bar c}_n^{a}|0,0,0\rangle =0, \ \ \ \ \ \hat{p}_c|0,0,0\rangle =0.
\ee
On this Hilbert space the oscillator Poisson brackets become the commutators
\be
[\hat{c}_m^A, \hat{c}_n^{B\dagger}] = \eta^{AB} \delta_{m,n}, \ \ \ \ \ \  [\hat{\bar{c}}_m^A, \hat{\bar{c}}_n^{B\dagger}] = \eta^{AB} \delta_{m,n}\ \ \ \ \ \ [\hat{x}_c^A,\hat{p}_{cB}] = i \delta^A_B,
\ee

There is an operator ordering ambiguity in the physical Hamiltonian, but not in the spatial diffeomorphism constraint.  Introducing an arbitrary constant $\alpha_0$ as the ordering ambiguity,   the complete quantization problem is
\be
\left[\frac{\alpha'}{2} \hat{p}_A^2 +   \sum_{n>0}n
\left( \hat{c}_{nA}^\dagger \hat{c}_n^A    +  \hat{\bar{c}}_n^{A\dagger}  \hat{\bar{c}}_{nA}   \right)  - 2\alpha_0 \right]|\Psi\rangle = E |\Psi\rangle,
\ee
 subject to the condition
 \be
 \left(\hat{L}_n - \hat{\bar{L}}_n^\dagger\right) |\Psi\rangle = 0. \label{qdiffeo}
 \ee
In the zero component of the last equation there is no operator ordering ambiguity ( $\alpha_0$ cancels in the difference), therefore we have the  level matching condition
\be
\left(\hat{L}_0 - \hat{\bar{L}}_0^\dagger\right) |\Psi\rangle = \sum_{n>0} n \left( \hat{c}_{nA}^\dagger \hat{c}_n^A -   \hat{\bar c}_{nA}^\dagger \hat{\bar c}_n^A     \right) | \Psi\rangle = 0. \label{levelm}
\ee
Defining $\displaystyle \hat{N} = \sum_{n>0} n \hat{c}_{nA}^\dagger \hat{c}_n^A $, and similarly for $\hat{\bar{N}}$, the eigenvalue equation becomes
 \be
 \left[\frac{\alpha'}{2} \hat{p}_A^2 + 2\hat{N}  - 2 \alpha_0 \right]|\Psi\rangle = E |\Psi\rangle,
 \ee
 after using the level matching condition (\ref{levelm}).

In conventional string theory, the physical states are annihilated by the Hamiltonian {\it constraint}.  The latter is of exactly the same form as our physical Hamiltonian.  Thus  conventional string theory  is the $E=0$ sector.  For our theory, by contrast, $E$ is in principle any number up to satisfying the eigenvalue equation, which gives the mass spectrum condition
\be
\frac{\alpha'}{2}  M^2 = 2N - (2\alpha_0 + E).
\ee
This is identical in form to that obtained in the usual bosonic string. The constant $\alpha_0$ and $E$ differ in their origin: the former is  a part of the definition of the Hamiltonian operator, and $E$ is an energy eigenvalue. It is of course possible to absorb the ordering ambiguity constant into the eigenvalue, and work with $\tilde{E} = E+2\alpha_0$.  Either way the  quantization problem is prescribed by the highest weight states given by
\be
\hat{L}_0 |\Psi\rangle = (E/2+ \alpha_0) |\Psi\rangle, \label{highwt}
\ee
 with the same for $\hat{\bar{L}}_0$.  The $E=0$ sector corresponds to the conformal field theory arising from the parameter values $\alpha_0=1, D=26$. The remaining sectors are not CFTs.

 Solution of the diffeomorphism constraint is obtained by requiring (as in usual string theory)  that
 \be
 \hat{L}_n |\Psi\rangle =0, \ \ \ \  \hat{\bar L}_n  |\Psi\rangle =0, \ \ \ n>0,
 \ee
 which gives weak imposition of the remaining constraints, ie.
 \be
 \langle \Psi| L_n |\Psi \rangle =0,  \ \ \ \  n\ne 0.
 \ee
 We note that, apart from $n=0$ part of it which gives level matching,  this solution of the  diffeomorphism conditions is ``strong'' in the sense that the theory we are considering requires only (\ref{qdiffeo}), ie. that the expectation value in the last equation be the same non-zero constant for both $L_n$ and $\bar{L}_n$. However such a solution appears not to be possible, partly for the same reason that strong  imposition of the constraint cannot be solved in the Fock quantization.

 The algebra of the $L_n$ and $\bar{L}_n$ operators gives an anomaly for each sector. However, since this  anomaly is the same for both,  it cancels in the spatial diffeomorphism generators  $\hat{l}_n = \hat{L}_n - \hat{\bar{L}}_n$. Thus  the diffeomorphism algebra is anomaly free,   and the expectation values of its generators vanishes on physical states.

 At this stage we have diagonalized the Hamiltonian and solved the constraints, while maintaining manifest target space Poincare invariance. Since there is no conformal invariance in the classical theory, there is no need to impose this symmetry in the quantum theory.  What remains is the question of negative norm states.

 This problem is solved in the same way as in the covariant quantization of the string (ie. without gauge fixing): it is known that negative norm states are removed in the non-critical string in $d<26$ for $a<1$ where $a$ is the parameter in the action of  $L_0$, ie. $(\hat{L}_0-a)|\Psi\rangle =0$ \cite{GSW}. In the present case we have $a= E/2+\alpha_0$, so the condition for removing negative norm states is
 \be
 d<26,  \ \ \ \ \ \ \ E/2 +\alpha_0<1.
 \ee

 This completes the discussion of the quantization in dust time gauge. We close with the following observations: (i) the quantum theory  contains the bosonic string as a special case, obtained by choosing $d=26$ and $ E/2+ \alpha_0=1$; this is  the conformally invariant sector of the theory. (ii) For fixed $\alpha_0$, which is part of the definition of the Hamiltonian operator, we can have any value of $E$ up to the constraint for removing negative norm states given above. Each value of $E$ specifies a highest weight representation specified by eqn. (\ref{highwt}), and each such representation fixes a spacetime time particle picture.  Thus the Hilbert space of the theory may be written as the tensor sum
  \be
 {\cal H} = \oplus_E  \  {\cal H}_E 
 \ee
where ${\cal H}_E$ (with $E/2 +\alpha_0<1$)  is the eigenvalue $E$ component in sum. The physical picture is specified by selecting an $E$ for the ``state of the universe" in this modified bosonic string, which then comes with a fixed particle spectrum.  This is a reasonable viewpoint also from  a mathematical perspective because each $E$ sector is superselected: there are no operators that connect different $E$ sectors. In this sense the full Hilbert space may be viewed as one describing a ``multiverse,'' where each element in the sum represents a ``universe" in the sense of a fixed particle spectrum. 
 
 For instance for $E + \alpha_0/2 =0$, which satisfies the  positive norm condition, we have
 \be
 \frac{\alpha'}{2}  M^2 = 2N.
 \ee
Since $N>0$ there are no tachyons in $d<26$ for this choice of $E$. A different value of $E$ gives a different mass formula and physical states. 

\section{Conclusions and Discussion}

We have constructed a  modification of the closed bosonic string by coupling the worldsheet degrees of freedom to a dust field with  timelike gradient. The resulting theory is not conformally invariant, but nevertheless resembles the conventional  bosonic string in some respects.  The essential new feature is that the Hamiltonian constraint of  the latter  becomes the physical Hamiltonian of our theory, and it is this feature that gives rise to the main differences  at the quantum level.

We provided a complete  quantization and showed that the theory has a number of interesting features. Foremost amongst these are that (i)  any target space dimensions $d\le 26$  is permitted (ii) a mass spectrum closely resembling the string one is recovered, (iii) in the zero eigenvalue sector the usual conformally invariant string is recovered if $d=26$, and (iv) each value of $E$ specifies a ``universe'' with a fixed particle spectrum. The theory contains in principle all such ``multiverse'' sectors, but we would pick the value of $E$ that characterizes  our observed particle spectrum.  From a mathematical perspective, removing conformal invariance by the addition of the dust field provides a very natural time gauge,  and renders all highest weight representations as physically allowable, in the sense of absence of anomalies.

 The quantization also shows that the theory with dust may be viewed as providing a classical analog of a non-critical string theory, defined by the restrictions $d<26$ and $a<1$ (the eigenvalue of $\hat{L}_0$). The low energy limit cannot of course be recovered via a beta function calculation, since there is no need to impose conformal invariance. However an alternative route to determining the low energy regime is to consider the evolution of states (eg. the graviton) via the Heisenberg equations of motion. Consider for example  the expectation value of the graviton operator
 \be
  g^{AB}(t,k) = \ _{sc}\langle \psi_k | \hat{c}^{(A\dagger}_1(t)\hat{\bar{c}}^{B)\dagger}_1(t) |\psi_k \rangle_{sc}
 \ee
 where the operators depend on dust time $t$,  and parameters in a suitable semiclassical state $|\psi \rangle_{sc}$  (eg. like a coherent Glauber state of the electromagnetic field) with momentum $k$. Then the Heisenberg equations of motion would give the evolution equations for this expectation value, which in a suitable $\hbar\rightarrow 0$ limit would give the classical low energy theory.

It may be argued that the new world sheet field we introduce is not ``fundamental'' in the same way that  the world sheet metric, tensor fields on the target space, or gauge charges at the ends of an open string might be. This of course depends on what one views as fundamental fields from a purely reductionist point of view. Our approach is operational in the sense of model building;  if new ingredients provide interesting and possibly useful physical and mathematical perspectives, then it is  worth pursuing regardless of metaphysical questions about what is fundamental.

This work is the first exploration of this type of model, and it opens up at least  a few directions for further exploration.  The most obvious ones are the open string and D-branes, and beyond this, possible generalizations to include supersymmetry. Also of interest are non-Fock quantizations such as the polymer quantization method as applied to the scalar field  \cite{ALS, mah,hk}.  However of most interest is the question of whether such models, or variants thereof, can serve as unified theories of gravity and the other interactions in four space-time dimensions, without the need of unverified extra dimensions.

\section*{Acknowledgments }

The authors wish to thank Thea Gegenberg, Tim Koslowski, Gabor Kunstatter and Haitao Liu for helpful conversations, and  acknowledge financial support of NSERC.

\end{document}